\documentclass[aps,prb]{revtex4}
\usepackage{amsmath}
\usepackage{graphicx}

\begin{document}

\title{Contribution of the  acoustic waves to near-field  heat transfer}

\author{A.I. Volokitin$^{*}$}

\affiliation{
Department of Physics, Samara State Technical University, Samara, 443100, Russia}

\begin{abstract}
Calculations of  the radiative and phonon heat transfer between metals in an extreme near field in presence of electrostatic potential difference are given. Potential difference leads to a  coupling  between the radiation field and acoustic waves in solid, as a result of which the heat flux between two gold plates associated with $p$ -polarized electromagnetic waves increases by many orders of magnitude as the potential difference varies from 0 to 10V. The radiative heat transfer is compared with  the phonon heat transfer associated with the electrostatic  and van der Waals interactions between the surface displacements.  For large potential difference and small distances the radiative heat transfer is reduced to the electrostatic phonon heat transfer. A particular case of surface acoustic waves - Rayleigh waves is studied in details. Conditions are obtained for the existence of surface phonon polaritons associated with the interaction of Rayleigh waves with an electromagnetic field. The surface Rayleigh and bulk acoustic waves  can give contributions of the same order. The obtained results   can be used to control heat fluxes at the nanoscale using  the potential difference and to create coherent radiation sources based on  the properties of the Rayleigh waves
\end{abstract}
\maketitle


\vskip 5mm

\section{Introduction}
Radiative heat transfer at the nanoscale level is of fundamental importance \cite{Polder1971PRB,Pendry1999JPCM,Greffet2005SSR,Volokitin2001PRB,Volokitin2007RMP,Volokitin2017Book} and promises to find  application in a variety of
technologies ranging from thermophotovoltaic energy converters \cite{Jacob2014APL,Greffet2006JAP,BenAbdalakh2013SciRep,Chen2003APL,Park2008JQSRT,Svetovoy2014PRAp},  non-invasive thermal imaging \cite{Wilde2006Nature} to thermomagnetic information recording and processing \cite{Challener2009NatPhot,Biehs2014PRL,Biehs2013APL,Joulain2015APL} and nanolithography \cite{Pendry1999JPCM}.   Based on the fluctuation electrodynamics developed by Rytov \cite{Rytov1953,Rytov1967,Rytov1987}, it was theoretically predicted
\cite{Polder1971PRB,Pendry1999JPCM,Greffet2005SSR,Volokitin2001PRB,Volokitin2007RMP}  and experimentally confirmed  \cite{Shen2009NanoLett,Greffet2009NatPhot,Reddy2015AIP,Kittel2017NatCommun,Reddy2017NatCommun,Reddy2015Nature}, that the radiative heat flux between two bodies with different temperatures in the near field (when the distance between the bodies   $d<\lambda_T=c\hbar/k_BT$: at room temperature $\lambda_T\sim 10\mu$m) can be by many orders of magnitude larger than the limit, which is established by Planck's law for blackbody radiation.  With the development of new experimental techniques over the past decade, super-Planckian heat transfer has been observed for vacuum gaps between bodies in the interval from hundreds of nanometers to several {\AA}ngstr\"{o}ms    \cite{Reddy2015AIP,Kittel2017NatCommun,Reddy2017NatCommun,Reddy2015Nature}.
Generally, the results of these measurements turned out to be in good agreement with the predictions based on the fluctuation electrodynamics for a wide range of materials and geometries.

Despite of the important progress achieved  in understanding radiative heat transfer,  there are still remain significant unresolved problems in understanding
heat transfer between bodies in an extreme near field (gap size $<10$ nm)\cite{Kittel2017NatCommun,Reddy2017NatCommun}. In particular, the heat transfer observed in Ref.\cite{Kittel2017NatCommun} for the separations from 1 to 10nm is orders of magnitude larger than the predictions of conventional Rytov theory and  its distance
dependence is not well understood. These discrepancies stimulate  the search of the alternative channels of the heat transfer that can be activated in an extreme near-field. One of the obvious channel is related with  Òphonon tunnelingÓ  which stimulated active research of the phonon heat transfer in this region \cite{Persson2011JPCM,Budaev2011APL,Joulain2014PRB,Pendry2016PRB,Pendry2017Z.Nat,Sellan2012PRB,Esfarjani2015NCom,KapitzRes2016PRB,Roy2010PRL,Meltaus2010PRL,Henkel2019JOSA,Volokitin2019JETPLett}. Most papers  consider phonon tunneling mediated by the van der Waals interaction. In Ref.\cite{Pendry2016PRB} it was suggested another mechanism for phonon tunneling based on  electrostatic potential differences between surfaces.  For radiative heat transfer, phonon tunneling was considered for polar dielectrics whose surfaces can support  surface phonon polaritons\cite{Volokitin2017Book,Volokitin2004PRB}. Metal  surfaces can support surface plasmon polaritons and they also can give important contribution to the radiative heat transfer\cite{Boriskina2015Photonics,Mahan2017PRB} but for good conductors, like gold, the frequencies of the surface plasmon polaritons are too high and they do not contribute significantly to the radiative heat transfer. Giant enhancement of the radiative heat transfer due to resonant phonon tunneling between adsorbate vibrational modes occurs between metal surfaces covered by adsorbed layer with ions\cite{Volokitin2017Book,Volokitin2003JETPLett,Volokitin2004PRB}. The contribution  from the acoustic waves are usually not considered in  the conventional theory of the radiative heat transfer because they are  optically nonactive. However, in Ref.\cite{Volokitin2019JETPLett} it was shown that the acoustic waves can be activated by the electrostatic potential difference.  The potential difference induces a 
surface charge density on the surface of the metals, as a result of which thermal fluctuations of the surface displacements give an additional contribution to the fluctuating electromagnetic 
field and, as a result, to the radiative heat transfer. For the charged surfaces all the acoustic waves in the frequency range $0<\omega<c_s/d<k_BT/\hbar$ can contribute to the heat transfer where $c_s$ is the sound velocity and $d$ is the separation between the surfaces. In an experiment electrostatic potential difference exists when STM is used for measurement of the heat transfer\cite{Kittel2017NatCommun,Reddy2017NatCommun}. Only the contribution  from the bulk acoustic waves was considered in Ref. 
\cite{Volokitin2019JETPLett} The surfaces can support Rayleigh acoustic waves that propagate near the surface of the medium  and do not penetrate deep into it \cite{Landau1970ThElasticity}. Rayleigh waves are one of the varieties of surface acoustic waves that are widely used in processing high-frequency signals, delay lines, sensors, and, more recently, for manipulating particles in microchannels. However, the contribution of the Rayleigh  waves to the near-field radiative heat 
transfer was not considered so far 
because there is no coupling between the thermal radiation and acoustic waves.   In this article the contribution from the Rayleigh waves is 
calculated and compared with bulk contribution. It is shown that these contributions can be of the same order. The radiative heat transfer is compared with the  
phonon heat transfer which is due to the electrostatic and van der Waals interactions between fluctuating surface  displacements. For the electrostatic and van der Waals phonon heat transfer the contributions from the Rayleigh and bulk acoustic waves can be also of the same order what does not agree with the results 
of Ref.\cite{Pendry2016PRB}.

\section{Theory}
\subsection{Radiative heat transfer}

Consider two identical metal plates, which are separated from each other by a vacuum gap with a thickness  $d$. Plate 1 has a surface that coincides with the $xy$ plane  and is located for $z <0$, while parallel plate 2   is located for $z> d$. A potential difference $ \varphi $ is applied between the surfaces, which induces a surface charge density
\begin{equation}
\sigma_0=\frac{\varphi}{4\pi d}=\frac{E_0}{4\pi },
\label{s0}
\end{equation}
where $E_0$ is the electric field in the gap between plates.
The radiative heat transfer is associated with the fluctuating electromagnetic field created by
thermal fluctuations of the charge  and current densities inside the bodies. In the case of a charged surface, thermal fluctuations of the surface displacement will also contribute to the fluctuating electromagnetic field and radiative heat transfer. The heat flux between two surfaces separated by a vacuum gap $d$ due to  evanescent (non-radiative) electromagnetic waves (for which $q>\omega/c$)  is determined by the formula
\cite{Volokitin2001PRB,Volokitin2007RMP,Volokitin2017Book}
\begin{equation}
J^{rad} =\frac{1}{\pi^2}\int_0^\infty d\omega\left[\Pi_1(\omega)-\Pi_2(\omega)\right]\int_0^{\infty} k_zdk_ze^{-2k_zd}
\left[\frac{
\mathrm{Im}R_{1p}(\omega,q)\mathrm{Im}R_{2p}(\omega,q) }{\mid 1-e^{-2
k_z d}R_{1p}(\omega,q)R_{2p}(\omega,q)\mid ^2}+(p\rightarrow s)\right],
\label{Heat}
\end{equation}
where
\[
\Pi_i(\omega)=\frac{\hbar \omega}{e^{\hbar\omega/k_BT_i}-1},
\]
 $R_p$ and  $R_s$  are the reflection amplitudes  for $p$- and  $s$-polarized electromagnetic waves, $k_z=\sqrt{q^2-(\omega/c)^2}$, $\mathbf{q}$ is the component of the wave vector parallel to the surface.  In the case of charged surfaces, the reflection amplitudes 
for them will no 
longer be determined by the Fresnel formulas, since  the interaction of a charged 
surface with 
an electric field produces a mechanical stress, which will give rise to surface 
polarization due to the displacement of the surface. In the presence of a surface 
dipole moment, the boundary conditions for the electric field are determined by equations \cite{Volokitin2019JETPLett,Langreth1989PRB}
\begin{equation}
E_z^+=\varepsilon(\omega)E_z^-,
\label{bcd1}
\end{equation}
\begin{equation}
E_q^+-E_q^-=-4\pi iq p_z=-4\pi iq\sigma_0^2ME_z^+,
\label{bcd2}
\end{equation}
where $E_z^{\pm}=E_z(z=\pm0)$, $E_q$ is  the component of the electric field  along $\hat{q}$-vector, $p_z= \sigma_0u$ is the normal component of the surface dipole moment, $u=M\sigma^{mech}$ is the surface displacement under the action of the mechanical stress $\sigma^{mech}=\sigma_0E_z^+$, $M$  is the mechanical 
susceptibility that determines the surface displacement 
under the action of mechanical stress. 
The reflection amplitude 
 for $\textit{p}$ -polarized electromagnetic waves obtained using boundary conditions (\ref{bcd1}) and (\ref{bcd2}) is given by 
\cite{Volokitin2019JETPLett}
\begin{equation}
R_p=\frac{i\varepsilon k_z  -k_z^{\prime} +
4\pi iq^2\sigma_0^2M\varepsilon}{i\varepsilon k_z  +
k_z^{\prime} -4\pi iq^2\sigma_0^2M\varepsilon},
\label{rcp}
\end{equation}
where $k^{\prime}_z=\sqrt{\varepsilon(\omega/c)^2-q^2}=\sqrt{(\varepsilon-1)(\omega/c)^2-k_z^2}$, 
$\varepsilon$ is the dielectric function of the medium.

\subsection{Phonon heat transfer}

 For the electrostatic and van der Waals interactions between surfaces 1 and 2  the stresses that act on surfaces 1 and 2 due to their  displacements  are determined by 
equations\cite{Volokitin2019JETPLett} (see also Appendixes \ref{A} and \ref{B} for detailed  derivation)
\begin{equation}
\sigma_1=au_{1}-bu_{2},
\label{sigma1}
\end{equation}
\begin{equation}
\sigma_2=au_{2}-bu_{1},
\label{sigma2}
\end{equation}
\begin{equation}
a=\frac{H}{2\pi d^4} + \frac{E_0^2}{4\pi}\frac{q\left(e^{qd}+e^{-qd}\right)}{e^{qd}-e^{-qd}},
\label{avdw}
\end{equation}
\begin{equation}
b=\frac{H}{4\pi}\frac{q^2K_2(qd)}{d^2} + \frac{E_0^2}{2\pi}\frac{q}{e^{qd}-e^{-qd}},
\label{bvdw}
\end{equation}
where the first and second terms are due to the van der Waals and 
electrostatic interactions, respectively, $H$ is the Hamaker constant, 
$K_2(x)$ is the modified Bessel function of the second kind and second order. 
The   surface displacements  due to thermal and quantum fluctuations are determined by \cite{Volokitin2019JETPLett,Persson2011JPCM}
\begin{equation}
u_1= u_1^f+M_1(au_1-bu_2),
\label{eqg}
\end{equation}
\begin{equation}
u_2= u_2^f+M_2(au_2-bu_1),
\label{eqd}
\end{equation}
where according to the fluctuation-dissipation theorem, the spectral density of fluctuations of the surface displacements is determined by  \cite{LandauStatisticalPhysics}
\begin{equation}
\langle|u_i^f|^2\rangle = \hbar \mathrm{Im}M_i(\omega,q)\coth\frac{\hbar\omega}{2k_BT_i},
\label{fdt}
\end{equation}
where $M_i$ is the mechanical susceptibility for surface $i$. 

Due to the electrostatic and van der Waals interactions, the fluctuating displacements of the surfaces will create fluctuating stresses acting on the surfaces, resulting in a heat flux between the surfaces that is determined
by  \cite{Volokitin2019JETPLett}
\begin{equation}
J^{ph}=\frac{1}{\pi^2}\int_0^\infty d\omega\left[\Pi_1(\omega)-\Pi_2(\omega)\right]\int_0^\infty dq q
\frac{b^2\mathrm{Im}M_1 \mathrm{Im}M_2}{\mid (1-aM_1)(1-aM_2)-
b^2M_1M_2\mid^2}.
\label{heatel}
\end{equation}

\section{Numerical results}
\subsection{Contribution from bulk acoustic waves}

In a elastic continuum model
 \cite{Persson2001JPCM}
\begin{equation}
M=\frac{i}{\rho c_t^2}\left(\frac{\omega}{c_t}\right)^2\frac{p_l(q,\omega)}{S(q,\omega)},
\label{M}
\end{equation}
where
\[
S(q,\omega)=\left[\left(\frac{\omega}{c_t}\right)^2-2q^2\right]^2+4q^2p_tp_l,
\]
\[
p_t=\left[\left(\frac{\omega}{c_t}\right)^2-q^2+i0\right]^{1/2}, \,\,p_l=\left[\left(\frac{\omega}{c_l}\right)^2-q^2+i0\right]^{1/2},
\]
where   $\rho$,  $c_l$, and
$c_t$  is the density of the medium, the velocity of the longitudinal and transverse acoustic waves, respectively. For gold: $c_l=3240$ms$^{-1}$, $c_t=1200$ms$^{-1}$, $\rho=1.9280\times10^4$kgm$^{-3}$,
 $H=34.7\times 10^{-20}$J (see Ref.\cite{Hamaker2015}).
The dielectric function of gold \cite{Chapuis2008PRB}
\begin{equation}
\varepsilon=1-\frac{\omega_p^2}{\omega^2+i\omega\nu},
\end{equation}
where  $\omega_p=1.71\times10^{16}$s$^{-1}$, $\nu=4.05\times10^{13}$s$^{-1}$.

\begin{figure}
\includegraphics[width=1.0\textwidth]{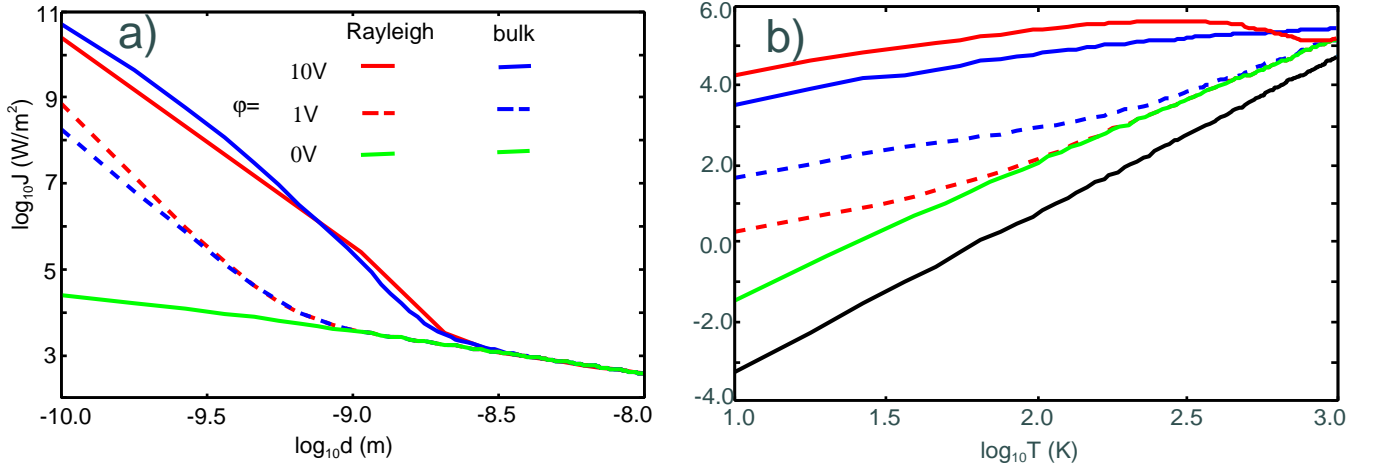}
\caption{Dependence of the radiative heat flux from $p$-polarized electromagnetic waves between two plates
of gold on (a) the distance at $T=300$K and (b) temperature at $d=1$nm  for various contributions.  Red solid and dashed lines for the radiative heat flux with  contributions from Rayleigh waves at the potential difference 10 and 1V, respectively. Blue solid and dashed  lines include the contributions from bulk acoustic waves at the potential difference 10, 1V, respectively.  Green lines show the results without potential difference when there is no coupling between radiation field and the acoustic waves. Black line on (b) is for the radiative flux associated with the black body radiation.  \label{Ral.Rad.Dist}}
\end{figure} 

\begin{figure}
\includegraphics[width=1.0\textwidth]{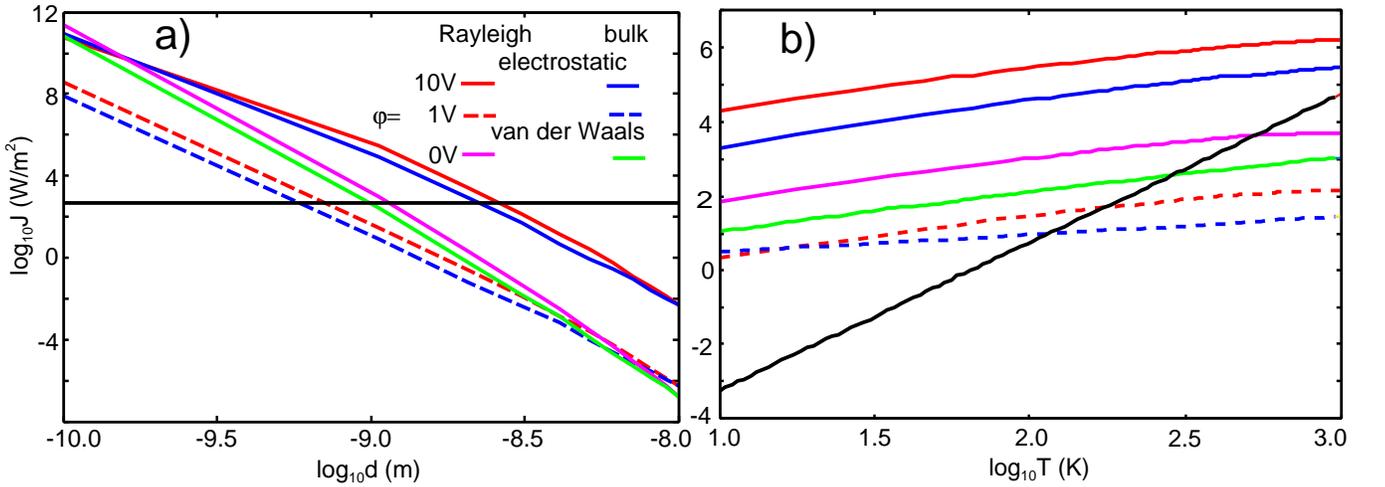}
\caption{The same as on Fig. \ref{Ral.Rad.Dist} but for the phonon heat transfer.   Red solid
and dashed lines for the contributions of  the Rayleigh waves for the electrostatic interaction  at the potential difference 10 and
1V, respectively. Blue solid and dashed lines for the contributions of the  bulk acoustic waves at 10
and V = 1V. Pink and green lines for contributions of the Rayleigh waves and bulk acoustic waves associated with
the van der Waals interaction.  \label{Ral.VdW.El.Dist}}
\end{figure} 

\begin{figure}
\includegraphics[width=0.5\textwidth]{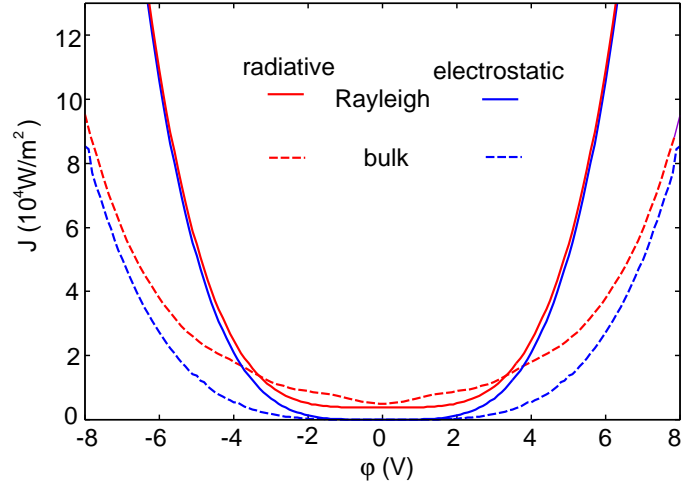}
\caption{Dependence of the heat flux  between two plates of gold on the potential difference  between plates  
 for different mechanisms. Red and blue lines for the radiative and electrostatic phonon heat transfer mechanisms. Solid and dashed lines show the contributions from the Rayleigh and bulk acoustic waves, respectively. $d=1$nm  \label{Ral.Rad.El.Voltage}}
\end{figure}

\begin{figure}
\includegraphics[width=0.5\textwidth]{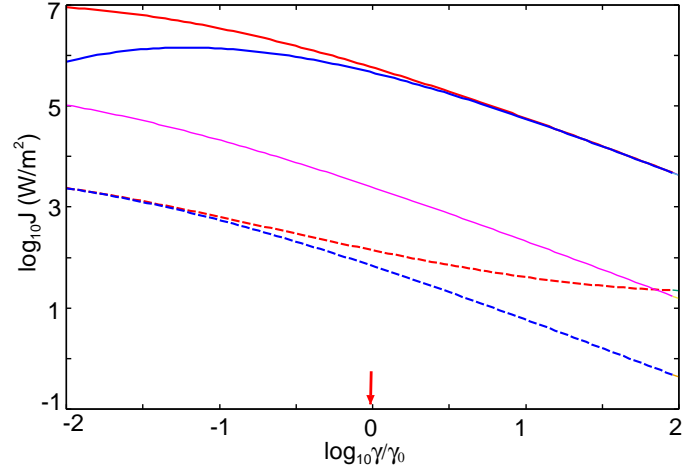}
\caption{Dependence of the contributions of the Rayleigh waves to the heat flux  between two plates of gold on the relative damping constant $\gamma/\gamma_s$ of these  waves  for different mechanisms at $d=1$nm and $T=300$K. Red and  blue  lines are for the radiative and  electrostatic  phonon heat transfer mechanisms at the potential difference 10V (solid lines) and 1V (dashed lines), respectively. Pink line is for van der Waals phonon heat transfer mechanism. Red arrow shows $\gamma=\gamma_0$ where $\gamma_s$ is the default damping constant given by Eq. (\ref{gammas}). 
\label{Damping}}
\end{figure}

The reflection coefficient (\ref{rcp}) has poles, which are determined by the equation
\begin{equation}
\mathrm{Re}\left(\varepsilon -i\frac{k_z^{\prime}}{q}  -4\pi q\sigma_0^2M\varepsilon\right)=0.
\label{polus}
\end{equation}
For high frequencies, when  $\varepsilon \rightarrow 1-\omega_p^2/\omega^2$, $k_z^{\prime}\approx iq$ and $4\pi q\sigma_0^2M\ll 1$, Eq. (\ref{polus}) is reduced to  $\varepsilon+1=0$ which determines the frequency of the surface plasmon polarotons $\omega_{sp}=\omega_p/\sqrt{2}$. 
In an extreme near field ($d\ll c/\omega|\varepsilon|$, $ik_z\approx k_z^{\prime}\approx iq$) 
for good conductors 
($|\varepsilon|\gg $) and $4\pi q\sigma_0^2M\gg  1/|\varepsilon|$ (\ref{rcp}) is reduced to
\begin{equation}
R_p=\frac{1   +4\pi q\sigma_0^2M}{1   -4\pi q\sigma_0^2M}.
\label{apprrcp}
\end{equation}
For  
$1/|\varepsilon|\ll4\pi q\sigma_0^2M\ll  1$, (\ref{Heat}) and (\ref{heatel}) are reduced to
\begin{equation}
J^{rad} =\frac{1}{\pi^2}\int_0^\infty d\omega\left[\Pi_1(\omega)-\Pi_2(\omega)\right]
\left[\frac{E_0^4}{4\pi^2}\int_0^{\infty} dq q^3\left(\frac{e^{-qd}}{1-e^{-2qd}}\right)^2
\mathrm{Im}M_{1}(\omega,q)\mathrm{Im}M_{2}(\omega,q)\right],
\label{C2}
\end{equation}
\begin{equation}
J^{ph}=\frac{1}{\pi^2}\int_0^\infty d\omega\left[\Pi_1(\omega)-\Pi_2(\omega)\right]
\int_0^\infty dq q
\left(\frac{H}{4\pi}\frac{q^2K_2(qd)}{d^2} + \frac{E_0^2}{2\pi}\frac{q}{e^{qd}-e^{-qd}} \right)^2
\mathrm{Im}M_1 \mathrm{Im}M_2.
\label{C3}
\end{equation} 
For $qd\gg 1$: $K_2(qd)\approx \sqrt{\pi/2qd}e^{-2qd}$ (see Ref.\cite{Handbook}), thus 
the $q$-integration in 
(\ref{C3}) is restricted by $q<1/d$, from where follows that 
the electrostatic contribution to the phonon heat transfer dominates for $d>H/2\varphi^2$ when
(\ref{C3}) is reduced to (\ref{C2}). The underlying physics of this result is that, with a large potential difference and small distances, the electrostatic field created by the fluctuating dipole moment of the charged surfaces mainly determines the fluctuating electromagnetic field. In this case, the radiative heat transfer is reduced to the electrostatic phonon heat transfer.
For two identical bodies, introducing new variables $q=(\omega/c_t)x$ and $\omega=\omega_ty$ where $\omega_t=c_t/d$, at $\hbar\omega_t/k_BT\ll 1$ Eq.(\ref{C2})  can be written in the form
\begin{equation}
J^{rad} =\frac{E_0^4\omega_t^3k_BT}{96\pi^2\rho^2c_t^6}\int_0^1dx\left[
\mathrm{Re}\frac{\sqrt{(c_t/c_l)^2-x^2}}{(1-2x^2)^2+4x^2\sqrt{1-x^2}\sqrt{(c_t/c_l)^2-x^2}}\right]^2.
\label{jradtemp}
\end{equation}
 
\subsection{Contribution from Rayleigh waves}

From formula (\ref{M}) it follows that the mechanical susceptibility $M$ has poles, when
\begin{equation}
S(q,\omega)=\left[\left(\frac{\omega}{c_t}\right)^2-2q^2\right]^2+4q^2p_tp_l=0.
\label{S}
\end{equation}
These poles determine the frequency of Rayleigh surface waves that propagate near 
the surface of the medium  and do not penetrate deep into it \cite{Landau1970ThElasticity}. Eq. (\ref{S}) has a solution for $\omega=\omega_s=c_sq=\xi c_tq$, where  $\omega_s$ and $c_s$ are the frequency
and the propagation velocity of the Rayleigh surface waves. The number $\xi$ depends only on $c_t/c_l$ which  varies for different materials   from $1/\sqrt{2}$ to 0 wherein $\xi$ varies from 0.874 to 0.955. For gold $c_t/c_l=0.37$ and $\xi=0.94$. Near the pole at $\omega\approx \omega_s$ the mechanical susceptibility can be written in the form
\begin{equation}
M=-\frac {C}{\rho c_t(\omega-c_sq+i\gamma)},
\label{mpol}
\end{equation}
where
\[
C=\frac{c_{st}(1-c_{sl}^2)\sqrt{1-c_{sl}^2}}{4[\sqrt{1-c_{st}^2}\sqrt{1-c_{sl}^2}(c_{st}^2-2)+1+c_{tl}^2-2c_{sl}^2]},
\]
$c_{ij}=c_i/c_j$,  for gold $C=0.36$, $\gamma$ is the damping constant for surface wave. If $\gamma=i0$, as it is assumed in Eq. ($\ref{M}$), then the Rayleigh  waves do not contribute to the heat transfer. As it was shown in Ref.\cite{Pendry2016PRB} for the damping constant of the  Rayleigh  waves can be used estimate
\begin{equation}
\gamma_0=0.17\omega_s=1.97\times10^2q \,s^{-1}.
\label{gammas}
\end{equation}
From (\ref{apprrcp}) it follows that for a charged surface in the low-frequency region a resonance may occur, which is determined by the equation
\begin{equation}
1 -4\pi q\sigma_0^2\mathrm{Re}M=0.
\label{lfrres}
\end{equation}
Equation (\ref{lfrres}) determines the frequencies of the surface phonon polaritons 
associated with the interaction of Rayleigh waves
 with electromagnetic field.
Using (\ref{mpol}) in (\ref{lfrres}) we get equation for the frequency of the surface phonon polaritons
\begin{equation}
1+\frac{c_0q(\omega-c_sq)}{(\omega-c_sq)^2+\gamma^2}=0,
\label{spol}
\end{equation}
where
\begin{equation}
c_0=\frac{4\pi\sigma_0^2C}{\rho c_t}=\frac{ C}{4\pi\rho c_t}\left(\frac{\varphi}{ d}\right)^2.
\end{equation}
Eq. (\ref{spol}) has solution at the condition $c_0q/\gamma=c_0/0.17\xi c_t>1$. However, for gold this condition can be fulfilled for non-realistic values of the potential difference and distances. In realistic case, when $c_0/c_t\ll 1$, the contribution of the Rayleigh  
waves to the heat transfer according to Eq. (\ref{C2}) is determined by 
\begin{equation}
J^{rad}_R \approx\left(\frac{CE_0^2}{2\pi^2\rho c_t}\right)^2\int_0^\infty d\omega\frac{\hbar \omega}{\mathrm{exp}(\hbar\omega/k_BT)-1}
\int_0^{\infty} dq q^3\left(\frac{e^{-qd}}{1-e^{-2q d}}\right)^2
\left[\frac{\gamma}{(\omega-\omega_s)^2+\gamma^2}\right]^2\approx\frac{k_BTc_0^2}{d^3c_t}.
\label{RadHeatSW}
\end{equation}
At zero potential difference, i.e. for noncharged surfaces, the contribution of $p$-waves to the heat transfer is determined by
\begin{equation}
J^{rad}_p\approx 0.2\frac{(k_BT)^3}{\hbar^2cd}\left(\frac{k_BT\nu}{\hbar\omega_p^2}\right)^{1/2}.
\label{Heatp}
\end{equation}
At $d=0.5$nm: $J^{rad}_R(\varphi=10$V$)\approx 5\times 10^7Ish$W/m$^2$ 
and $J^{rad}_p(\varphi=0$V$) \approx 2\times 10^4$W/m$^2$. For the electrostatic 
and van der Waals interactions the contribution of the Rayleigh  waves to the phonon 
heat transfer according to Eq. (\ref{C3}) is determined by 
\begin{equation}
J^{ph}_R \approx\frac{1}{\pi^2}\int_0^\infty d\omega\frac{\hbar \omega}{\mathrm{exp}(\hbar\omega/k_BT)-1}
\int_0^{\infty} dq q^3\left(\frac{CHqK_2(qd)}{4\pi\rho c_td^2}+
\frac{CE_0^2}{2\pi\rho c_t}\frac{e^{-qd}}{1-e^{-2q d}}\right)^2
\left[\frac{\gamma}{(\omega-\omega_s)^2+\gamma^2}\right]^2.
\label{PhHeatSW}
\end{equation}
 The contribution from the electrostatic interaction dominates for $d>H/\varphi^2$ (see discussion after Eq.(\ref{C3})), when $J^{ph}_{R}\approx J^{rad}_R$.
The contribution from the van der Waals interaction dominates for $d<H/\varphi^2$ when
\begin{equation}
J^{ph}_R \approx \frac{1}{2\pi^2}\int_0^\infty d\omega\frac{\hbar \omega}{\mathrm{exp}(\hbar\omega/k_BT)-1}\int_0^{\infty} qdqK_2^2(qd)
\left[\frac{c_{vdw}q\gamma}{(\omega-\omega_s)^2+\gamma^2}\right]^2\approx 1.9\frac{k_BTc_{vdw}^2}{d^3c_t},
\label{PhHeatSW}
\end{equation}
where
\[
c_{vdw}=\frac{CH}{4\pi\rho c_td^3},
\]
and it was used the value of the integral 
\[
\int_0^{\infty}x^4K_2^2(x)dx=6.06.
\]
With distance, the contributions due to the van der Waals  and  electrostatic interactions decay as $d^{-9}$ and $d^{-7}$, respectively. 
Figs. \ref{Ral.Rad.Dist}, \ref{Ral.VdW.El.Dist} and \ref{Ral.Rad.El.Voltage} compare the contributions from the Rayleigh   and volume acoustic waves to the radiative and phonon heat transfer. It follows from the calculations that these contributions are of  the same order and have almost the same distance dependence, which does 
not agree with the results of Ref.\cite{Pendry2016PRB}. On Fig.\ref{Ral.Rad.El.Voltage} 
the difference between the red and blue 
lines decreases when the potential difference increases what confirm theoretical prediction that at large potential difference and small separation the radiative heat transfer is reduced to the  electrostatic phonon heat transfer. Fig. \ref{Damping} shows the dependence of the contributions to  the heat fluxes due to the Rayleigh waves  on the  damping constant of these waves.

\section{Conclusion} 
The calculations of heat transfer between two plates of gold in an extreme near field are performed at the presence of the potential difference between the plates. The presence of a potential difference leads to a coupling between
the  radiation field and the acoustic waves, due to which the radiative heat transfer increases by many orders of magnitude when the potential difference varies from 0 to 10V. Radiative 
heat transfer was compared with the phonon heat transfer due to  the electrostatic  and van der Waals interactions between the surface displacements.  For large potential difference and small distances the radiative heat transfer is reduced to the electrostatic phonon heat transfer. The presence of a potential difference leads to a contributions from Rayleigh surface waves, which are of the same order and have almost the same distance dependence as the contribution from bulk acoustic waves. Similar to the  surface electromagnetic
waves that are arising  due to the interaction of the electromagnetic field with optical waves, Rayleigh waves on a charged surface can also
create a thermal electromagnetic field with spatial and temporal coherence.  The near-field properties of the thermal electromagnetic field in the presence
of surface electromagnetic waves were reviewed in Ref.\cite{Joulain2005SSR}.  The heat flux  between a gold coated near-field scanning thermal microscope tip and a planar gold sample  in an extreme near field at nanometer distances of 0.2-7nm  in the presence of the electrostatic potential difference of $\sim$1V between the tip and the sample was studied  in Ref.\cite{Kittel2017NatCommun,Reddy2017NatCommun}. It was found that the the experimental results can not be explained by the conventional theory of the radiative heat transfer  based on Rytov theory.  In this article it was shown that the very large contribution, which is not included in the conventional Rytov theory,  can be due to  the fluctuating dipole moment induced on the surface by the potential difference.  On the base of the presented theory it can be proposed that anomalously large nano-scale heat transfer between metals can be explained by the potential difference.  In addition, based on the data obtained, it can be assumed that the fluctuating dipole moment associated with the surface double layer of the metal can also make a large contribution to heat transfer. Further research is needed for a detailed comparison of theory with experiment. In the extreme near field the electron tunnelling can also contribute to the heat transfer. However, this contribution is unlikely (see discussion in Ref. \cite{Kittel2017NatCommun,Henkel2019JOSA}).  According to the experimental dada \cite{Kittel2017NatCommun,Reddy2017NatCommun} for  the distances above 1nm the tunnelling current is negligible. The tunnelling current grows exponentially for the distances below 0.5nm,  while the heat flux has the power dependence in this distance range, thus the electron tunnelling can be excluded from the heat transfer mechanism even in the extreme near field.  The obtained results can be used to control heat fluxes at the nanoscale using the potential difference. 

\appendix

\section{van der Waals interaction between surfaces \label{A}}

The van der Waals interaction energy between two semi-infinite media 1 and 2, whose surfaces
 have small displacements $u_i (\mathbf{x})$ with respect to the planes at $z = 0$ and $z = d$, can be written in the form
\begin{equation}
W =-\frac{H}{\pi^2}\int d^2\mathbf{x}_1\int d^2\mathbf{x}_2\int_{-\infty}^{u_1(\mathbf{x}_1)}dz_1\int^{\infty}_{u_2(\mathbf{x}_2)}dz_2\frac{1}{[(\mathbf{x}_1-\mathbf{x}_2)^2+
(d+z_2-z_1)^2]^3},
\label{uvdw}
\end{equation}
where $H$ is the Hamaker constant. Expanding $W$ in a series in powers of $u_i$ to terms that are quadratic in $u_i$ inclusive, we get
\[
W=-\frac{AH}{12\pi d^2}-\frac{H}{6\pi d^3}\left[\int d^2\mathbf{x}_1u_1(\mathbf{x}_1)-\int d^2\mathbf{x}_2u_2(\mathbf{x}_2)\right]+
\]
\begin{equation}
+\frac{H}{\pi^2 }\int d^2\mathbf{x}_1\int d^2\mathbf{x}_2\frac{u_1(\mathbf{x}_1)u_2(\mathbf{x}_2)}{[(\mathbf{x}_1-\mathbf{x}_2)^2+
d^2]^3}-\frac{H}{4\pi d^4}\left[\int  d^2\mathbf{x}_1u_1^2(\mathbf{x}_1)+\int  d^2\mathbf{x}_2u_2^2(\mathbf{x}_2)\right]+...
\end{equation}
The stresses that act on the surfaces 1 and 2  when they are displaced are determined by the equations
\begin{equation}
\sigma_1=-\frac{\delta U}{\delta u_1(\mathbf{x})}=-\frac{H}{\pi^2 }\int d^2\mathbf{x}_2\frac{u_2(\mathbf{x}_2)}{[(\mathbf{x}-\mathbf{x}_2)^2+
d^2]^3}+\frac{H}{2\pi d^4}u_1(\mathbf{x}),
\end{equation}
\begin{equation}
\sigma_2=-\frac{\delta U}{\delta u_2(\mathbf{x})}=-\frac{H}{\pi^2 }\int d^2\mathbf{x}_1\frac{u_1(\mathbf{x}_1)}{[(\mathbf{x}_1-\mathbf{x})^2+
d^2]^3}+\frac{H}{2\pi d^4}u_2(\mathbf{x}),
\end{equation}
where displacement-independent terms, which do not contribute to heat transfer, were omitted. Using
representation of displacements as a Fourier integral
\begin{equation}
u_i(\mathbf{x}_i)=\int \frac{d^2\mathbf{q}}{(2\pi)^2}u_ie^{i\mathbf{q}\cdot \mathbf{x}},
\end{equation}
we get Eqs. (\ref{sigma1}) and (\ref{sigma2}),   in which \cite{Pendry2016PRB,Pendry2017Z.Nat}
\begin{equation}
a=\frac{H}{2\pi d^4},
\label{avdw}
\end{equation}
\begin{equation}
b=\frac{H}{4\pi}\frac{q^2K_2(qd)}{d^2},
\label{bvdw}
\end{equation}
where  $K_2(x)$ is the modified Bessel function of the second kind and second order. In deriving (\ref{bvdw}), it was used the value of the integral \cite{Handbook}
\begin{equation}
\int d^2\mathbf{x}_1\frac{u_1(\mathbf{x}_1)}{[(\mathbf{x}_1-\mathbf{x})^2+
d^2]^3}=\int \frac{d^2\mathbf{q}}{(2\pi)^2}u_ie^{i\mathbf{q}\cdot \mathbf{x}}G(q),
\end{equation}
where
\begin{equation}
G(q)=\int d^2\mathbf{x}\frac{e^{i\mathbf{q}\cdot\mathbf{x}}}{(\rho^2+
d^2)^3}=2\pi\int_0^{\infty}\frac{J_0(q\rho)\rho d\rho}{(\rho^2+d^2)^3}=\frac{\pi q^2K_2(qd)}{4d^2}.
\end{equation}

\section{Electrostatic interaction between surfaces \label{B}}

For metal surfaces, the potential difference $\varphi$ between them  induces a surface charge density
\begin{equation}
\sigma_0=\frac{\varphi}{4\pi d}.
\label{s0}
\end{equation}
The interaction of closely spaced charged surfaces will be determined in the electrostatic limit, 
following the approach from Ref. \cite{Pendry2016PRB}. We assume that surface 1 at $z = 0$ is held at zero potential, and surface 2 at $z = d$ has a potential equal to $\varphi$. In the absence of displacements of the surfaces, the potential in the vacuum gap has the form
\begin{equation}
\varphi_0(z)=\frac{\varphi}{d}z.
\end{equation}
Displacements of the surfaces
\begin{equation}
u_i(\mathbf{x})=u_{i}e^{i\mathbf{q}\cdot \mathbf{x}}
\end{equation}
will lead to a change of the electrostatic field in the vacuum gap at $0 <z <d$, which can be described by potential 
\begin{equation}
\varphi(\mathbf{x},z)=\varphi_0(z)+\left(\nu_- e^{-qz}+\nu_+ e^{qz}\right)e^{i\mathbf{q}\cdot\mathbf{x}}.
\end{equation}
In the electrostatic limit, the potential of metal surfaces should remain unchanged when
the surfaces are displaced. From where follow the boundary conditions 
\begin{equation}
E_0u_{1}+\nu_- +\nu_+ =0,
\label{elbcg}
\end{equation}
\begin{equation}
E_0u_{1}+\nu_-e^{-qd}+\nu_+e^{qd}=0 ,
\label{elbcd}
\end{equation}
where $E_0=\varphi/d=4\pi \sigma_0$.
Eqs. (\ref{elbcg}) and (\ref{elbcd}) can be also obtained from the boundary condition (\ref{bcd2}). From (\ref{elbcg}) and (\ref{elbcd}) we get
\begin{equation}
\nu_+ =\frac{E_0u_{1}e^{-qd}}{e^{qd}-e^{-qd}}-\frac{E_0u_{2}}{e^{qd}-e^{-qd}},
\label{nu+}
\end{equation}
\begin{equation}
\nu_- =-\frac{E_0u_1e^{qd}}{e^{qd}-e^{-qd}}+\frac{E_0u_1}{e^{qd}-e^{-qd}}.
\label{nu-}
\end{equation}
The normal component of the electric field  at $0 <z <d$ has the form
\begin{equation}
E_z =-E_0-\frac{qE_0}{e^{qd}-e^{-qd}}\left[\left(e^{q(z-d)}+e^{q(d-z)}\right)u_{1}-\left(e^{qz}+e^{-qz}\right)u_{2}\right)e^{i\mathbf{q}\cdot \mathbf{x}}.
\label{efg}
\end{equation}
Using the Maxwell stress tensor, one can find the mechanical stresses that act on the
surfaces 1 and 2 at their displacements
\begin{equation}
\sigma_1=\frac{|E_z(z=0)|^2}{8\pi}-\frac{E_0^2}{8\pi}\approx \frac{E_0^2}{8\pi}\left[\frac{2q\left(e^{qd}+e^{-qd}\right)}{e^{qd}-e^{-qd}}u_{1}e^{i\mathbf{q}\cdot \mathbf{x}}
-\frac{4q}{e^{qd}-e^{-qd}}u_{2}e^{i\mathbf{q}\cdot \mathbf{x}}\right],
\label{tildes11}
\end{equation}
\begin{equation}
\sigma_2=-\frac{|E_z(z=d)|^2}{8\pi}+\frac{E_0^2}{8\pi}\approx \frac{E_0^2}{8\pi}\left[\frac{2q\left(e^{qd}+e^{-qd}\right)}{e^{qd}-e^{-qd}}u_{2}e^{i\mathbf{q}\cdot \mathbf{x}}-\frac{4q}{e^{qd}-e^{-qd}}u_{1}e^{i\mathbf{q}\cdot \mathbf{x}}\right]
\label{tildes21}
\end{equation}
 From Eqs. (\ref{tildes11})  and (\ref{tildes21}) follow Eqs. (\ref{sigma1}) and (\ref{sigma2}), where
\begin{equation}
a=\frac{E_0^2}{4\pi}\frac{q\left(e^{qd}+e^{-qd}\right)}{e^{qd}-e^{-qd}},
\label{a}
\end{equation}
\begin{equation}
b=\frac{E_0^2}{2\pi}\frac{q}{e^{qd}-e^{-qd}}.
\label{b}
\end{equation}
From (\ref{a}), (\ref{b}), (\ref{avdw}) and  (\ref{bvdw}) follow, that $a\approx b$ for $qd\ll 1$. In this case, the Òspring modelÓ becomes valid,
which was considered in Refs.\cite{Persson2011JPCM,Joulain2014PRB,Budaev2011APL}. However, quantitatively the Òspring modelÓ is valid only for
unrealistic short distances \cite{Pendry2016PRB}.

\vskip 0.5cm

The reported study was funded by RFBR according to the research project N\textsuperscript{\underline{o}} 19-02-00453
\vskip 0.5cm

$^*$alevolokitin@yandex.ru

\end{document}